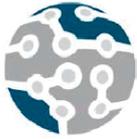
XV Jornadas de Ingeniería Telemática.
JITEL 2021.
Universidad de A Coruña.

*Actas de las XV Jornadas de Ingeniería Telemática (JITEL 2021), A Coruña (España), 27-29 de octubre de 2021.*# Confiabilidad en la capa de transporte para la red de sensores antártica

Adrià Mallorquí, Agustín Zaballos, Alan Briones, Guiomar Corral
Grupo de Investigación en Internet Technologies and Storage (GRITS)
La Salle – Universitat Ramon Llull
Quatre Camins 30, Barcelona 08022
adria.mallorqui@salle.url.edu, agustin.zaballos@salle.url.edu, alan.briones@salle.url.edu, guiomar.corral@salle.url.edu**El proyecto de investigación SHETLAND-NET aspira a desarrollar un servicio de telemetría de *Internet of Things* (IoT) en la Antártida mediante la interconexión de *Wireless Sensor Networks* (WSN) a través de radioenlaces *Near Vertical Incidence Skywave* (NVIS) que conforman una *Long Fat Network* (LFN). Esta arquitectura presenta algunas propiedades típicas de las denominadas *challenging networks*, requiriendo una evaluación de la viabilidad de la solución propuesta y un análisis de qué protocolo de transporte puede aportar una mayor confiabilidad para este caso de uso. Para este propósito, se define y presenta un modelo de confiabilidad heterogéneo basado en capas. A través de extensivas simulaciones se valida el modelo, se comparan distintos protocolos de transporte y se evalúa la confiabilidad del sistema.**

*Palabras Clave*- Protocolo de transporte, confiabilidad, Antártida, Long Fat Network, IoT, NVIS## I. Introducción

Es bien sabido que en la Antártida se llevan a cabo múltiples estudios científicos de distintos ámbitos de investigación [1]. Durante las campañas de investigación, los investigadores se establecen temporalmente en bases antárticas, normalmente ubicadas en las zonas periféricas del continente antártico. Uno de los mayores retos de la Antártida es la falta de sistemas de telecomunicaciones convencionales [1], hecho que dificulta el despliegue de *Wireless Sensor Networks* (WSN), reduciendo las posibilidades de añadir mejoras a los estudios actuales (p. ej. mediante la comunicación entre dos bases remotas o automatización en la recolección de datos).

Para superar estas dificultades, el proyecto de investigación SHETLAND-NET continúa nuestro estudio del uso de radioenlaces de Alta Frecuencia (HF) usando la técnica *Near Vertical Incidence Skywave* (NVIS), proporcionando comunicaciones de bajo consumo en la Antártida. La señal transmitida rebota en la ionosfera, consiguiendo un *long backhaul link* con un área de cobertura de 250 km de radio y un ancho de banda de pocos kbps [2], [3]. Las redes con este tipo de enlaces se pueden clasificar como *Long Fat Networks* (LFN), caracterizadas por contener enlaces con un *Bandwidth Delay Product* (BDP) mayor a 12.500 bytes [4].

La tecnología NVIS se puede utilizar para interconectar bases científicas remotas, y nuestro objetivo es acabar desplegando una red *Internet of Things* (IoT) que interconecte WSN remotas [2]. Este despliegue en el campo lo llevaremos a cabo durante la siguiente campaña antártica. Sin embargo, las comunicaciones NVIS pueden ser propensas a errores según el estado de la ionosfera. Se pueden presentar situaciones típicas de las *challenging networks* [5], como por ejemplo conectividad intermitente, desconexiones puntuales extremo a extremo y tasas de error variables, que pueden degradar el rendimiento del servicio IoT en caso de emplearse una arquitectura de protocolo TCP/IP estándar.

Es necesario que, antes de la fase de despliegue, podamos estudiar e intentar anticipar la confiabilidad esperada del sistema de telemetría IoT que queremos desarrollar. De esta forma podríamos prever los posibles problemas de confiabilidad que podrían surgir y escoger qué contramedidas aplicar respectivamente.

Para este trabajo, nos centraremos en el caso de uso de la automatización de las estaciones *Ground Terrestrial Network-Permafrost* (GTN-P) [6]. Cada una de estas estaciones GTN-P se equipa con una placa Arduino que recoge 32 valores de distintos sensores cada hora. Estos valores deben ser enviados hasta el centro de control situado en una base científica. El Arduino de cada estación mandará estos valores hacia un concentrador Raspberry Pi 3B+ a través de comunicaciones *Long Range* (LoRa) en la red de acceso. Esta Raspberry actuará como *gateway*, reenviando los datos recibidos hacia el centro de control a través de los enlaces NVIS (red *backbone*).

La fiabilidad de los enlaces NVIS depende mucho del estado de la ionosfera y de la actividad solar de forma que, durante la noche, no es posible mandar datos empleando la

This work is licensed under a Creative Commons 4.0 International License (CC BY-NC-ND 4.0)*63*

misma frecuencia de transmisión que durante el día (prácticamente todos se perderían). Por esta razón, aplicamos una técnica de envío oportunista propio de las *Delay Tolerant Networks* (DTN) para enviar todos los datos recolectados durante la noche cuando el enlace NVIS pasa a estar disponible por la mañana. Cada concentrador debería haber recolectado 13 sets distintos de valores por cada estación GTN-P durante la noche. Se espera que durante este momento de envío oportunista la red se congestione debido a la cantidad de datos trasmitidos. Nuestro proyecto requiere que, de media, un mínimo de 9 de los 13 sets de datos de cada estación (aproximadamente un 70%) lleguen al centro de control correctamente. Creemos que es necesario evaluar qué protocolo de transporte se utilizará para este envío de una gran cantidad de datos [4], ya que puede influir en el rendimiento y la confiabilidad del servicio, especialmente en situaciones de congestión. Por este motivo, queremos estudiar y comparar el uso de distintos protocolos de transporte sobre la LFN modelando el escenario y el caso de uso en el simulador Riverbed Modeler.

El resto del artículo se estructura de la siguiente forma. En la sección II se describe el trabajo relacionado en protocolos de transporte y confiabilidad de los sistemas. En la sección III se presenta nuestra propuesta de modelo para medir y evaluar la confiabilidad de un sistema. En la sección IV se detallan los test y simulaciones ejecutados. En la sección V se discuten los resultados obtenidos. Finalmente, en la sección VI se comentan las conclusiones.

## II. Trabajo relacionado

### A. Protocolos de transporte

El rendimiento de los protocolos de transporte ha sido un tema de discusión y desarrollo desde que Internet fue concebido [7]. Los protocolos de transporte tradicionales como *Transmission Control Protocol* (TCP) sufren un bajo rendimiento en ciertos tipos de redes, entre las cuales se encuentran las LFNs. El concepto de LFN y sus efectos sobre el rendimiento de TCP fueron definidos en la *Request for Comments* (RFC) 1072, que fue posteriormente actualizada por la RFC 1323 y finalmente la RFC 7323. Algunas variantes de TCP y otros protocolos de transporte desarrollados durante los últimos años han mejorado el rendimiento de la transmisión para LFN [8]. Sin embargo, estos interpretan que las pérdidas de paquetes siempre son originadas por una congestión en la red, reduciendo así la tasa de transmisión o ventana de congestión. Esta asunción no es cierta para redes inalámbricas, donde los paquetes también se pueden perder debido a la propia naturaleza del medio (p. ej. el *fading*, la movilidad o las interferencias) [9]. Reducir la ventana de congestión en estas situaciones también degrada el rendimiento de la transmisión, consiguiendo un *throughput* menor. Por este motivo, existen protocolos de transporte que implementan mecanismos para discernir entre pérdidas originadas por congestión y pérdidas por canal, para así solo reducir la ventana de congestión en el primer caso y aumentar el rendimiento del envío [10].

TCP CUBIC (RFC 8312) [7] es el protocolo de transporte más utilizado actualmente, ya que es la variante de TCP utilizada por defecto en la mayoría de sistemas operativos. Además, otros protocolos modernos como TCP BBR, Copa, Indigo y Verus [10] son capaces de conseguir un buen rendimiento, tal y como se ha demostrado en los extensivos test realizados por la plataforma Pantheon [8] de la Universidad de Standford. En el presente artículo nos apoyaremos en nuestro trabajo previo: el *Adaptive and Aggresive Transport Protocol* (AATP) y su evolución, el *Enhanced AATP* (EAATP), el cual incorpora un mecanismo para diferenciar el motivo de una pérdida de paquetes y otro mecanismo de *fairness* para adaptar la tasa de envío según las circunstancias estimadas de la red [4], [10].

### B. Confiabilidad en Cyber Physical Systems

Un *Cyber Physical System* (CPS) se define como un sistema con capacidades físicas y computacionales integradas. Algunos ejemplos de CPS son los sistemas de control industrial, las redes eléctricas inteligentes y las WSNs, así como la mayoría de dispositivos que abarcan el IoT [11]. En la literatura se define la confiabilidad de un CPS, en términos generales, como la propiedad de un sistema para comportarse como se espera ante situaciones adversas [11]. Estas adversidades pueden venir derivadas de distintos motivos, como por ejemplo nodos defectuosos, errores bizantinos, comportamientos maliciosos y errores de red, entre otros. Por esta razón, se pueden encontrar distintos enfoques para medir y proporcionar confiabilidad que se basan en elementos dispares. Proponemos clasificar estos enfoques en las siguientes cuatro categorías, que serán la base para definir nuestro modelo a continuación:

*1) Confiabilidad de los datos*: se define como la posibilidad de poder confirmar la exactitud de un dato proporcionado por una fuente [12]. Existen varios métodos que pretenden detectar nodos defectuosos o fallos de lectura de valores. Por ejemplo, en [13] se utiliza una arquitectura de *fog computing* para detectar, filtrar y corregir datos anormales. En [14] se utiliza un sistema de detección de intrusión de datos para notificar datos erróneos provenientes de ataques maliciosos.

*2) Confiabilidad de la red*: se define como la probabilidad de que un paquete llegue a su destino a tiempo y sin ser alterado a pesar de las adversidades (p. ej. un error de enlace, la saturación del canal o ataques maliciosos, entre otros) [15]. La mejora de la confiabilidad de la red es un tema que se ha enfocado desde diferentes perspectivas como los protocolos de transporte, los protocolos de direccionamiento y control topológico [16] o las arquitecturas DTN [5].

*3) Confiabilidad social*: esta tendencia ha ganado atención desde la irrupción del concepto *Social Internet of Things* (SIoT) [17], [18]. Para la confiabilidad en SIoT, se utiliza la capacidad social de los nodos u objetos a la hora de establecer relaciones autónomamente para poder definir modelos de confianza y reputación que tengan en cuenta distintos factores evaluados por los propios nodos. En [19] se define un modelo subjetivo que considera factores como la capacidad computacional de los nodos, el tipo de relación entre ellos, el número total de transacciones, la credibilidad de un nodo o el *feedback* recibido de otros nodos, entre otros. En [20] se propone otro modelo que define unos parámetros de entrada como el beneficio esperado ante un éxito, las pérdidas esperadas ante un fallo, el coste esperado y la importancia del objetivo, entre otros. En [21] se define un sistema descentralizado de autogestión de la confianza basado en un sistema de *feedback* reputacional donde se preserva la privacidad de las partes participantes.

*4) Consenso*: representa el estado en el que todos los participantes de un mismo sistema distribuido acuerdan un mismo valor o respuesta [22]. Los protocolos de consenso se pueden dividir en dos grandes bloques: los consensos *proof-based* y los consensos bizantinos. Los primeros están más orientados a tecnologías *blockchain*, donde todos los participantes compiten por minar un bloque, y los más utilizados son *Proof-of-Work*, *Proof-of-Stake* y sus variantes [22]. El principal inconveniente de estos protocolos para IoT es que la mayoría de dispositivos tienen un hardware sencillo y una capacidad computacional reducida, dificultando así las tareas de minado [22]. El segundo bloque de protocolos de consenso es el más clásico, orientado a la detección de errores bizantinos. Estos suelen implementar mecanismos cooperativos de votación para llegar a un acuerdo en vez de hacerlo de forma competitiva, consiguiendo un menor consumo de recursos. El principal problema de estos mecanismos es la baja escalabilidad debido a la gran cantidad de mensajes que deben intercambiar los nodos participantes. Los protocolos más utilizados en este ámbito son *Practical Byzantine Fault Tolerance* (PBFT), RAFT, PaXoS y Ripple, aunque más variantes han surgido a lo largo de estos últimos años [22].

### III. MODELO DE CONFIABILIDAD

Después de nuestra revisión bibliográfica, todos los trabajos que se pueden encontrar en la literatura se centran en áreas específicas de la confiabilidad, pero ninguno de ellos incluye las cuatro categorías. Este hecho puede llevar a no interpretar correctamente las causas subyacentes del nivel de confiabilidad, de forma que unas contramedidas no idóneas, o incluso contraproducentes, se podrían aplicar si las interdependencias entre las distintas categorías de confiabilidad no se consideran. Por este motivo, encontramos la necesidad de diseñar un modelo propio que exprese y permita trabajar adecuadamente el nivel de confiabilidad de un sistema e incluya las cuatro categorías mencionadas anteriormente. Este modelo debe ayudarnos a anticipar e identificar los posibles puntos débiles de nuestro sistema de telemetría IoT.

Nuestro modelo propuesto para medir la confiabilidad y evaluar el rendimiento de un CPS (en nuestro caso, un grupo de WSN remotas interconectadas para proporcionar un servicio de telemetría IoT en la Antártida) se basa en cuatro capas. El modelo está caracterizado por 1) dos capas base (Capa de Confiabilidad del Dato y Capa de Confiabilidad de la Red), 2) dos capas de extensión (Capa de Confiabilidad Social y Capa de Consenso) que incluye funcionalidades opcionales, y 3) las interacciones entre ellas. Postulamos que cada capa se caracteriza por su definición (alcance), cómo se mide la confiabilidad de esa capa (métrica), y cómo se puede mejorar el valor de esta métrica (contramedidas).

#### A. Capa de Confiabilidad del Dato

Esta capa tiene como objetivo confirmar la exactitud de los datos obtenidos por la fuente. Proponemos medir la confiabilidad de esta capa con la métrica *Faulty Sensing Ratio* ($FSR$), definida en la Ec. 1 como la proporción entre el número de valores sensados erróneamente ($FSV$) y el número total de valores sensados ($TSV$) en mismo periodo de tiempo. Cuánto más bajo sea el $FSR$, mejor será la confiabilidad de los datos.

$$FSR = \frac{FSV}{TSV} \qquad (1)$$

Se pueden aplicar métodos autocorrectivos que intenten mitigar el efecto de datos anormales ($FSV$) en el nodo origen [13], [14]. Otros ejemplos son los *hashes*, *checksums*, y bits de paridad, entre otros (ver Fig. 1).

#### B. Capa de Confiabilidad de la Red

Esta capa es la responsable de asegurar que los paquetes lleguen a su destino a tiempo y sin ser alterados a pesar de las adversidades (p. ej. error de enlace o saturación del canal). Medimos la confiabilidad de esta capa con el *Packet Delivery Ratio* ($PDR$), definido en la Ec. 2 como el cociente entre el número total de paquetes correctamente recibidos por todos los nodos ($Pr$) y el número total de paquetes enviados por todos los nodos ($Ps$), durante el mismo periodo de tiempo. Cuanto más alto sea el $PDR$, mejor será la confiabilidad de la red.

$$PDR = \frac{Pr}{Ps} \qquad (2)$$

En esta capa se pueden utilizar técnicas de codificación de la transmisión [23] para incrementar la robustez de la señal transmitida. También se utilizan habitualmente protocolos de encaminamiento y *Quality of Service* (QoS) para encontrar el mejor camino hacia un destino mediante los cuales se cuantifica la calidad o el rendimiento de los enlaces de la red [16]. Los protocolos de transporte y mecanismos de control de congestión también pueden mejorar la confiabilidad de la red [10]. En el caso de las *challenging networks* se utilizan arquitecturas y protocolos *overlay* DTN, como por ejemplo el *Bundle Protocol* [5].

#### C. Capa de Confiabilidad Social

Esta capa es la responsable de beneficiarse de la capacidad de los nodos de establecer relaciones sociales autónomamente para mejorar la confianza entre ellos y escoger con más probabilidad los valores correctos. Medimos la confiabilidad de esta capa con la métrica *Successful Transaction Rate* ($STR$), calculada como la proporción entre el número de transacciones satisfactorias ($STR$) y el número total de transacciones ($TT$) en un mismo período de tiempo, tal y como se muestra en la Ec. 3. Una transacción $l$ se considera satisfactoria cuando un nodo $j$ espera recibir alguna información o dato $v$ del nodo $i$ antes de un tiempo máximo de recepción ($Trx_{max}$) y el valor recibido entra dentro de un intervalo esperado, de forma que el nodo $j$ proporciona un *feedback* positivo sobre el nodo $i$ ($f_{ij}^l = 1$). Cuanto mayor sea la $STR$, mejor será la confiabilidad social.

$$STR = \frac{ST}{TT} \quad (3)$$

Las contramedidas en esta capa suelen utilizar mecanismos de reputación para determinar en qué nodos confiar como proveedores y/o receptores de datos. Esta reputación se puede basar en experiencias previas, tanto por parte del nodo evaluador como de los nodos vecinos, que ayudan a construir una opinión sobre la confianza hacia otro nodo concreto [19], [21].

*D. Capa de Consenso*

Esta capa persigue alcanzar un estado donde todos los participantes de un grupo acuerden una misma respuesta o resultado conjunto (*General Agreement* o GA). Medimos la confiabilidad de esta capa con la métrica *Byzantine Node Tolerance* ($BNT$), definida como el cociente entre el número de nodos bizantinos ($Nb$) que se toleran en un grupo sin afectar al acuerdo llegado y el número total de nodos participantes en ese grupo ($Nt$), tal y como se muestra en la Ec. 4. Un nodo se considera bizantino si experimenta un fallo que le incapacite para comportarse como se espera o si no sigue el comportamiento esperado a propósito (malicioso). Cuanto mayor es la $BNT$, más probable es conseguir un GA correcto.

$$BNT = \frac{Nb}{Nt} \quad (4)$$

Se pueden utilizar multitud de mecanismos para obtener un GA descentralizado. Teóricamente, si el número de nodos bizantinos supera la mitad de los nodos totales participantes, cualquier mecanismo de consenso fallará al intentar conseguir un GA correcto [22]. El principal inconveniente de estos mecanismos es que los nodos participantes necesitan intercambiar un gran número de mensajes para conseguir el consenso, hecho que puede degradar el rendimiento de redes con poco ancho de banda.

*E. Relaciones entre las capas de consenso*

La Fig. 1 sintetiza los actores de nuestro modelo de confiabilidad. Los elementos azules forman parte de las capas base de nuestro modelo, mientras que los elementos naranjas forman parte de las capas de extensión. El objetivo fundamental es incrementar el $STR$ para conseguir una mayor confiabilidad, de forma que esta es la métrica principal que utilizaremos en nuestras simulaciones. Hay tres factores que directamente ayudan a aumentar el $STR$: 1) Mitigar/tolerar errores bizantinos; 2) reducir el $FSR$; y 3) aumentar el $PDR$. Estos factores los consideramos subobjetivos de nuestro modelo. Cada uno de estos subobjetivos se pueden conseguir mediante la implementación de sus respectivas contramedidas, que normalmente solo afectan a uno de los subobjetivos. Sin embargo, también detectamos dos acciones transversales que pueden afectar a más de un subobjetivo. Estas acciones transversales consisten en implementar las capas de extensión de nuestro modelo: la Capa de Confiabilidad Social y la Capa de Consenso.

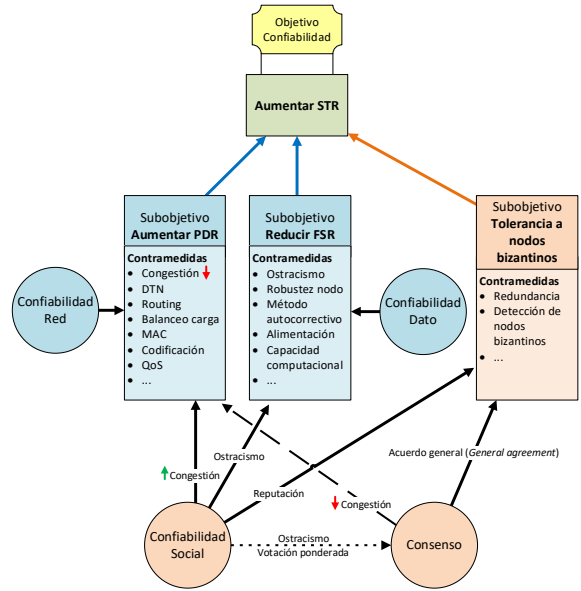

Fig. 1. Diagrama de relaciones del modelo de confiabilidad.

En la Fig. 1, las flechas con línea continua indican una afectación positiva, las flechas con línea discontinua indican una afectación negativa, y las flechas con línea punteada indican una afectación incierta. Por ejemplo: el uso de la confiabilidad social puede reducir la congestión de la red, gracias al ostracismo de los nodos con peor reputación, que pueden dejar de generar tráfico, e intercambiar solo los datos de los nodos con más reputación. Este hecho también ayuda a mitigar los errores bizantinos, ya que se priorizará el uso los nodos de más confianza (que tendrán menos probabilidades de experimentar un error bizantino). Pero, por otro lado, la implementación de un mecanismo de consenso ayuda a tolerar los errores bizantinos gracias al GA al que llegan los nodos participantes de un mismo grupo. Sin embargo, la Capa de Consenso puede afectar negativamente al $PDR$, ya que introduce una cantidad considerable de tráfico extra que puede llegar a congestionar la red.

IV. TEST Y SIMULACIONES

Con el objetivo de 1) anticipar qué problemas pueden ocurrir durante la campaña antártica, 2) decidir qué protocolo de transporte utilizar para nuestro servicio, y 3) tener unas expectativas sobre los resultados del despliegue más precisas, aplicamos nuestro modelo de confiabilidad para medir y evaluar el caso de uso propuesto. Con esta finalidad, el escenario del caso de uso se ha modelado y testeado en el simulador Riverbed Modeler. A continuación, se detalla cómo se han modelado cada uno de los actores de nuestro caso de uso.

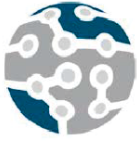

Tabla I
MODELO DE RED PARA LAS SIMULACIONES

| Parámetro | NVIS | LoRa |
|---|---|---|
| Banda de transmisión | 4.3 MHz | 868 MHz |
| Ancho de banda | 2.3 kHz | 125 kHz |
| *Bitrate* | 20 kbps | 5,47 kbps |
| Rango de cobertura | Hasta 250 km | Hasta 30 km |
| Disponibilidad diurna (6am-5pm) | 70%-100% | 100% (LoS), 2%-100% (No LoS) |
| Disponibilidad nocturna (5pm-6am) | 0% | 100% (LoS), 2%-100% (No LoS) |
| Tamaño máximo de *payload* | 242 bytes | 140 bytes |

Tabla II
PARÁMETROS DE LAS SIMULACIONES

| Parámetro | Valor |
|---|---|
| Número de rondas por test | 30 |
| Duración | 120 horas (5 días) |
| $Pb_0$ | [$1\times10^{-3}$, $2\times10^{-3}$, $4\times10^{-3}$, $8\times10^{-3}$, $1\times10^{-2}$, $2\times10^{-2}$, $4\times10^{-2}$, $8\times10^{-2}$, $1\times10^{-1}$] |
| $k$ | $5.7\times10^{-5}$ |
| Protocolo de transporte | [BBR, Copa, CUBIC, EAATP, Indigo, Verus] |
| Modo de redundancia | [Ninguno, Social, Consenso] |
| Número de *gateways* NVIS | 5 |
| Número de *clusters* GTN-P por *gateway* | [8, 16, 32, 64, 128, 256, 512, 1024, 2048, 4096] |
| Número de estaciones GTN-P redundantes por *cluster* | [1-10] |

En primer lugar, para el modelo de la red se han hecho dos modelos independientes para la red *backbone* (NVIS) y la red de acceso (LoRa), ya que usan tecnologías con características distintas. Estas redes se han caracterizado por separado tal y como muestra la Tabla I, partiendo de los resultados de [3] y [24] respectivamente. En segundo lugar, se han modelado los siguientes protocolos de transporte tal y como hicimos en nuestro trabajo previo [10]: BBR, Copa, CUBIC, EAATP, Indigo y Verus. Su rendimiento esperado se ha caracterizado según los resultados obtenidos en nuestro trabajo previo [10] y los test de la plataforma Pantheon [8].

En tercer lugar, es necesario modelar el comportamiento bizantino de los nodos. Según [25], la probabilidad que un nodo experimente un error bizantino, $Pb_0$, no es constante en el tiempo. De hecho, el incremento de esta probabilidad se puede asociar al desgaste del nodo, que también está relacionado con la descarga de la batería que lo alimenta. Siguiendo el modelo de [25], podemos asumir que el impacto del desgaste es lineal, como se define en la Ec. 5:

$$Pb(t) = Pb_0 + k \times t, \quad (5)$$

donde $Pb_0$ es la probabilidad de que un nodo experimente un error bizantino en el instante $t = 0$ y $k$ es el factor de desgaste. La probabilidad de que un nodo experimente un error bizantino incrementa a lo largo del tiempo hasta que la batería se vacía completamente en $t = t_d$. En este momento el nodo deja de responder y $Pb(t_d) = 1$. En las simulaciones se utilizan varios valores de $Pb_0$ para modelar el uso de distintos métodos correctivos.

A continuación, para el modelo de la confiabilidad social se ha utilizado una versión simplificada del modelo de confiabilidad objetiva de [19]. La simplificación para nuestro caso de uso sirve al escenario que se desplegará en la base antártica en el que todas las transacciones tienen la misma importancia, todos los nodos tienen la misma capacidad computacional, y el tipo de relación social entre todos los nodos es equivalente.

Finalmente, el protocolo de consenso se puede modelar sabiendo el tráfico extra que este añade a la red y el número de nodos bizantinos que este tolera ($Nb$). En nuestro caso, se ha escogido el protocolo PBFT [26] para implementar el consenso. El tráfico que este protocolo añade crece exponencialmente a medida que el número de nodos que participan en el mismo grupo de consenso ($Nt$) aumenta. Además, el número de nodos tolerados ($Nb$) se puede calcular según la Ec. 6:

$$Nb = \left\lfloor \frac{Nt-1}{3} \right\rfloor \quad (6)$$

La Tabla II sintetiza los parámetros de nuestras simulaciones.

V. DISCUSIÓN DE LOS RESULTADOS

Tras ejecutar todas las simulaciones, se ha calculado la media del valor de $STR$ obtenido por cada grupo de 30 test. Los resultados obtenidos tienen una desviación máxima del 0,68% con un intervalo de confianza del 99%. En nuestro caso de uso se pueden diferenciar tres modos de funcionamiento principales: el modo estándar, el modo de redundancia con la aplicación de la Capa de Confiabilidad Social, y el modo de redundancia con la aplicación de la Capa de Consenso. Por cada modo, se construye una matriz de N×M dimensiones con todas las combinaciones posibles de los parámetros de la simulación, donde N es el número de posibles combinaciones de *clusters* por *gateway* y estaciones GTN-P por *cluster* (100 en nuestro caso), y M es el número de distintos valores de $Pb_0$ posibles (9 en nuestro caso). Por cada punto de esta matriz y por cada protocolo de transporte se calcula el valor medio de $STR$ obtenido. Si se unen todos los valores de $STR$ calculados, podemos obtener una malla por cada protocolo de transporte. A esta malla la llamamos Malla de Confiabilidad. Las Fig. 2, 3 y 4 muestran la Malla de Confiabilidad para el modo estándar, el modo de redundancia con confiabilidad social y el modo de redundancia con consenso, respectivamente.

El eje "Probabilidad de error bizantino" tiene 9 puntos discretos, correspondientes a los valores $Pb_0$ de la Tabla II. El eje "Sensores redundantes × Número de *clusters*" tiene 100 puntos discretos, que son [$1\times2^N$, $2\times2^N$, ..., $10\times2^N$], donde N = [3, 4, ..., 12]. Estos valores corresponden a los que se muestran en la Tabla II, filas 9 y 10. Aunque en las etiquetas del eje solo se muestran los valores iniciales de cada intervalo [1×8, 2×8, ..., 10×8], dentro de cada intervalo se incrementa el número de *clusters* por *gateway* (p.ej. [1×8, 1×16, ..., 1×4096]).

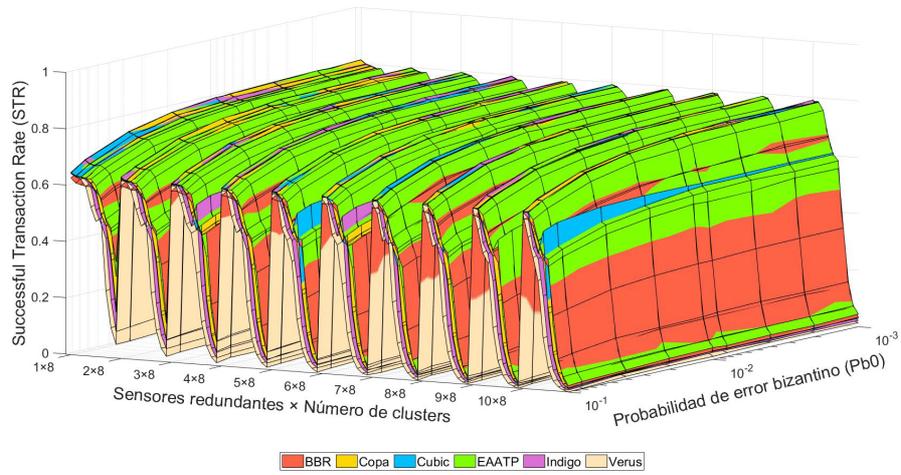

Fig. 2. Malla de confiabilidad (Estándar).

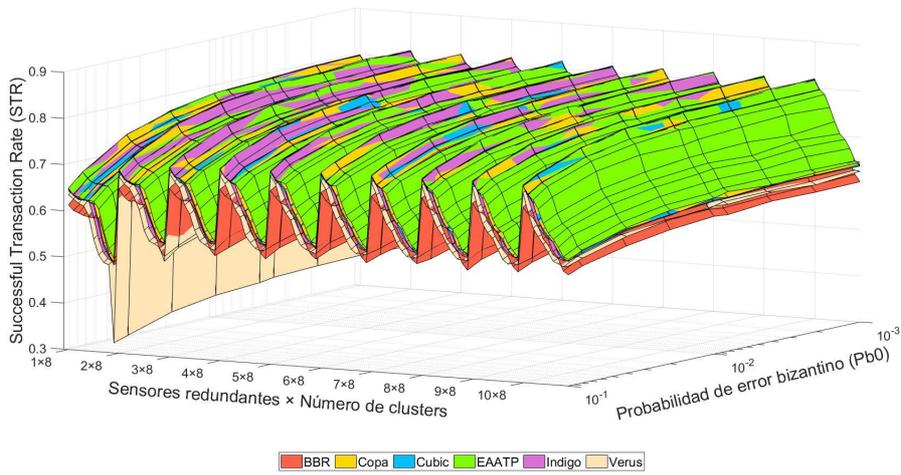

Fig. 3. Malla de confiabilidad (Social).

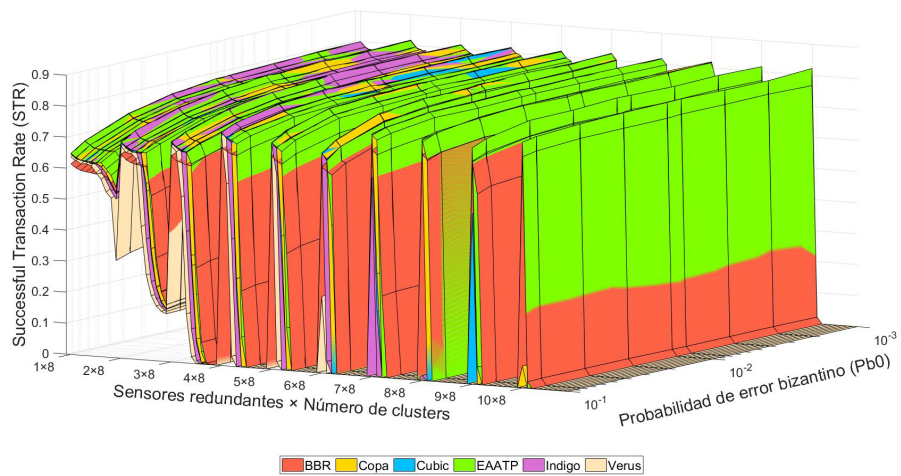

Fig. 4. Malla de confiabilidad (Consenso).

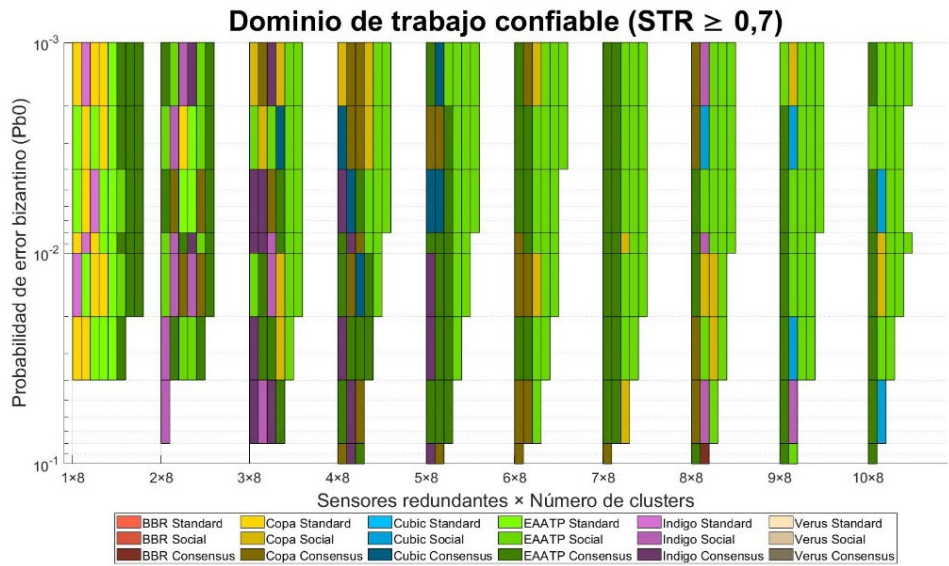

Fig. 5. Dominio de trabajo confiable para un $STR$ mínimo de 0,7. Se muestra la mejor opción para cada caso.

En las Fig. 2, 3 y 4 podemos ver, por un lado, que los niveles de confiabilidad obtenidos son similares por todos los protocolos de transporte estudiados cuando hay pocos nodos en la red y, por lo tanto, esta se encuentra poco cargada. Sin embargo, 1) los niveles de $STR$ obtenidos por BBR y Verus son ligeramente inferiores a los de sus competidores, y 2) Copa, Indigo y EAATP consiguen los valores de $STR$ más altos para una baja carga de la red, aunque el dominio del EAATP como el protocolo con más confiabilidad crece a medida que la sobrecarga de la red aumenta, cuando se generan más pérdidas por congestión.

Por otro lado, también observamos que el modo de redundancia con confiabilidad social (Fig. 3) es el más robusto, ya que los valores de $STR$ decrecen de una forma menos acentuada a media que la congestión y el número de nodos aumenta si lo comparamos con los otros casos (Fig. 2 y 4), manteniendo siempre unos valores superiores a 0,5. Además, se puede confirmar que en general tenemos más confiabilidad (mayores valores de $STR$) a medida que la $Pb_0$ decrece.

Nuestro modelo también se puede utilizar para visualizar el posible dominio de trabajo utilizable para implementar el servicio requerido, dado un valor mínimo de confiabilidad exigible. Nuestro caso de uso requiere un $STR$ mínimo de 0,7, de forma que un promedio de 9 de los 13 valores leídos por cada sensor lleguen correctamente al centro de control, y así cumplir con el objetivo de [6]. La Fig. 5 muestra el dominio de trabajo para nuestro servicio de telemetría y un $STR$ mínimo de 0,7. Por cada punto de la matriz, si no hay ninguna solución que consiga un $STR$ igual o superior al mínimo deseado, este se deja en blanco, significando que la red no puede satisfacer los requerimientos del servicio bajo esas condiciones. Contrariamente, si una o más soluciones consiguen un $STR$ igual o superior al mínimo deseado, ese punto se rellena con el color de la solución con un $STR$ mayor. Desde esta perspectiva se puede apreciar un claro dominio del EAATP como la solución más confiable para la mayoría de los casos. Concretamente, el EAATP puede llegar a mejorar hasta un 7% sus competidores (p.ej. en el caso "7×8" con $Pb_0=10^{-1}$, EAATP logra una $STR$ de 0,78 mientras que Verus obtiene 0,71), mientras que como máximo está un 0,5% por debajo cuando es superado por otro protocolo (p.ej., en el mismo caso que en el anterior, Copa logra una $STR$ de 0,785).

Concluimos que el dominio del EAATP se debe a sus mecanismos de *fairness* y diferenciación de pérdidas. Por un lado, el mecanismo de *fairness* es capaz de repartir el ancho de banda del enlace entre varios flujos EAATP sin que se generen pérdidas por congestión. Por otro lado, el mecanismo de diferenciación de pérdidas es capaz de detectar si un paquete se ha perdido debido a una congestión de la red o a un error de canal, de forma que no reduce la tasa de envío en el segundo caso y consigue un mayor rendimiento. Estos factores le dan una ventaja competitiva al EAATP, ya que en este caso de uso se utiliza una mecánica propia de las DTN para almacenar todos los valores en un nodo intermedio durante la noche y mandarlos en bloque cuando el canal se encuentra disponible, hecho que congestiona la red. Además, el protocolo EAATP está concebido para utilizar la máxima capacidad posible de un enlace, hecho que puede ser muy relevante en nuestro caso de uso ya que las redes disponibles tienen un ancho de banda reducido y es crucial poder utilizarlo al máximo.

## VI. CONCLUSIONES

Este artículo continúa el trabajo del proyecto de investigación SHETLAND, el cual persigue el objetivo de diseñar e implementar un conjunto de WSN remotas en la Antártida que se interconectan mediante el uso de enlaces NVIS. Nuestro trabajo se ha centrado en analizar y comparar la confiabilidad de diferentes protocolos de transporte para nuestro caso de uso, el cual ofrece un servicio de telemetría IoT. Debido a las características de la ionosfera, los enlaces NVIS no funcionan de la misma forma durante la noche, motivo por el cual los valores adquiridos durante este periodo se almacenan temporalmente en un nodo concentrador y se mandan en bloque cuando el canal se encuentra disponible, pudiendo congestionar la red. Para poder estudiar la viabilidad de esta arquitectura antes de implementar el servicio en la campaña antártica, y con el objetivo de comparar el rendimiento de varios protocolos de transporte, se ha definido un modelo para medir y evaluar la confiabilidad del sistema propuesto. Este modelo se compone de cuatro capas que pueden afectar a la principal métrica de confiabilidad, la *STR*, que mide el número de transacciones satisfactorias que llegan correctamente al centro del control y cuyo valor puede ser mejorado mediante la aplicación de contramedidas.

Se han analizado tres modos de funcionamiento y seis protocolos de transporte bajo distintas circunstancias con el simulador Riverbed Modeler. Los resultados muestran el dominio del protocolo EAATP como el más confiable para la mayoría de los casos (llegando a mejorar a sus competidores hasta un 7%), mientras que BBR y Verus son los menos confiables. Añadir redundancia de sensores y aplicar un método de confiabilidad social mejora la robustez del servicio, consiguiendo valores de *STR* más altos y que en ningún caso son inferiores a 0,5 incluso en situaciones de carga elevada. Contrariamente, aplicar un mecanismo de consenso mejora la confiabilidad del sistema en situaciones de pocos nodos, pero parece contraindicado cuando estos aumentan dada la sobrecarga de tráfico que introduce. En un futuro, se pretende evaluar el uso de distintas técnicas DTN para mejorar la confiabilidad del sistema ante situaciones donde la caída de los enlaces NVIS no sea predecible.